\documentclass{article}
\usepackage[a4paper, top=3cm, bottom=3cm, left=3cm, right=3cm]{geometry}

\usepackage[ruled,vlined]{algorithm2e}
\usepackage{subfig}
\usepackage[export]{adjustbox}
\usepackage{thmtools, thm-restate}
\usepackage{authblk}
\usepackage{hyperref}
\usepackage{amsmath}
\usepackage{amssymb}
\usepackage{amsthm}

\usepackage[dvipsnames]{xcolor}
\usepackage{stackengine}
\usepackage{graphicx}
\usepackage{tabularray}
\usepackage{tikz}
\usetikzlibrary{shapes,arrows}
\usepackage{enumitem}

\newcommand{\G}{\mathcal G}	
\newcommand{\I}{\mathcal I}	

\newtheorem{ex}{Example}

\newlength{\qedlengte}
\newcommand{\qedbox}{\rule{\qedlengte}{\qedlengte}}

\newenvironment{example}{\begin{ex}\rm}{\hfill\qedbox\end{ex}}

\def\codeif{\mbox{\upshape\textbf{if}}}
\def\codethen{\mbox{\upshape\textbf{then}}}

\def\codewhile{\mbox{\upshape\textbf{while}}}

\def\codereturn{\mbox{\upshape\textbf{return}}}

\newcommand{\comment}[1]{}

\newcommand{\Larrow}{\mathbin{\rotatebox[origin=c]{-20}{$\uparrow$}}}
\newcommand{\Uarrow}{\mathbin{\rotatebox[origin=c]{20}{$\downarrow$}}}
\newcommand{\Scolor}{ForestGreen}
\newcommand{\Xcolor}{Red}
\newcommand{\Acolor}{black}
\newcommand{\gradecolor}{Dandelion}
\newcommand{\SLarrow}[1]{${\color{\Scolor}\Larrow_{#1}}$}
\newcommand{\XLarrow}[1]{${\color{\Xcolor}\Larrow_{#1}}$}
\newcommand{\SUarrow}[1]{${\color{\Acolor}\Uarrow_{#1}}$}
\newcommand{\XUarrow}[1]{${\color{\Acolor}\Uarrow_{#1}}$}

\usepackage{MnSymbol}
\newcommand{\gradedXLarrow}[1]{${\color{\gradecolor}\Larrow_{#1}}$}

\usepackage{tikz}
\makeatletter
\newcommand*\redcircled[2][1.6]{\tikz[baseline=(char.base)]{
    \node[shape=circle, draw, red, very thick, inner sep=-1pt,
        minimum height={\f@size*#1},] (char) {#2};}}
\makeatother

\begin{document}

\title{Towards a visually interpretable analysis of Two-Phase Locking membership}

\author{Davide Martinenghi}

\affil{Politecnico di Milano, Italy,\quad \texttt{davide.martinenghi{@}polimi.it}
}
 
\date{}

\maketitle
\pagenumbering{arabic}
\pagestyle{plain}
\setcounter{page}{1}

\begin{abstract}
Two-phase locking (2PL) is a consolidated policy commonly adopted by Database Management Systems to enforce serializability of a schedule. 
While the policy is well understood, both in its standard and in the strict version, automatically deriving a suitable tabular/graphical analysis of schedules with respect to 2PL is far from trivial, and requires several technicalities that do not straightforwardly translate to visual cues.
In this paper, we delve into the details of the development of a tool for 2PL analysis.
\end{abstract}

\section{Introduction}\label{sec:introduction}

The Two-Phase Locking (2PL) policy is a fundamental concurrency control protocol used in database systems to ensure transaction serializability, a key property for maintaining consistency in multi-user environments. It was introduced in the 1970s in a seminal work on the theory of transaction management~\cite{DBLP:journals/cacm/EswarranGLT76}.
Before 2PL, database systems faced challenges with concurrent transactions causing anomalies such as lost updates, dirty reads, and uncommitted data dependencies. To address this, 2PL was designed to coordinate access to shared resources (e.g., data items) using locks, ensuring that the final execution order of transactions mimics a serial order.

2PL owes its name to the fact that it operates in two distinct phases for each transaction:
\begin{itemize}
\item \emph{Growing Phase}: A transaction can acquire locks (e.g., read or write locks) but cannot release any locks.
\item \emph{Shrinking Phase}: A transaction releases locks but cannot acquire new ones.
\end{itemize}
This structure ensures that transactions obey the principle of serializability, as all lock acquisitions and releases are well-ordered.

The simplicity and robustness of 2PL have led to its widespread adoption in traditional relational databases, such as IBM’s System R, PostgreSQL, and SQL Server. Over time, variations of 2PL have emerged to improve its performance and address practical issues~\cite{DBLP:books/mg/SKS20,DBLP:books/aw/BernsteinHG87}:

\begin{itemize}
\item \emph{Strict 2PL}: Requires transactions to hold all write locks until they commit or abort, simplifying recovery processes and ensuring durability.
\item \emph{Rigorous 2PL}: Mandates holding all locks (both read and write) until the transaction ends, offering even stricter guarantees.
\item Optimized Versions: Techniques like deadlock detection, lock escalation, and lock timeouts have been incorporated to mitigate performance bottlenecks.
\end{itemize}
In modern distributed systems, 2PL faces challenges due to its potential to cause deadlocks and increase transaction latency, particularly in high-concurrency scenarios. As a result, other concurrency control mechanisms, such as optimistic concurrency control and multi-version concurrency control, are often preferred in environments like cloud-native databases and NoSQL systems. Nevertheless, 2PL remains relevant in applications requiring strong consistency guarantees, such as financial systems and critical data processing.

Checking 2PL membership requires solving a system of inequalities deriving from the conditions imposed by the policy. However, solving such a system does not readily translate to visual interpretations corresponding to human intuition.
To this end, we discuss how to exploit a graph representation of the inequalities. In particular, the detection of minimal cycles in the graph plays a fundamental role in the graphical visualization of the analysis of 2PL membership.
Other visual cues, such as the representation of plateau points (between the growing and the shrinking phase) or a meaningful positioning of possible lock/unlock requests, require nontrivial bookkeeping.

In the rest of the paper, we describe the main principles underlying 2PL and illustrate the main steps required by schedule analysis and show the LaTeX rendering generated by a prototype tool that we have developed for this purpose.

\section{Checking and representing 2PL membership}

We adopt a simplified representation of operations on database resources that consists of read and write operations made by transactions.
An operation on a resource $x$ made by transaction $i$ is a triple that we represent as $r_i(x)$ if the operation is a read and $w_i(x)$ if it is a write.
A schedule is a sequence of such triples.
For convenience, we interpret as the \emph{time} of the triple the position in which the triple occurs in the schedule.
If we want to explicitly indicate the time $t$ of a triple, we write it as a superscript in square brackets as in $r_i^{[t]}(x)$ for read and $w_i^{[t]}(x)$ for write operations.

In a lock-based systems, an operation can only be executed if a proper lock is acquired by the executing transaction on the target resource: read operations require a shared lock ($SL$) or an exclusive lock ($XL$), while write operations require $XL$.
Some time after the operation, and before the end of the schedule, the lock must be released (indicated as $SU$ for shared unlock, i.e., releasing a shared lock and, similarly, $XU$ for the exclusive case).

The conditions imposed by the 2PL policy can be summarized as follows.

\begin{itemize}
	\item a proper lock request must precede an operation:
	\begin{itemize}
		\item $SL$ must precede the first read operation for a resource (skipped if the read operation is preceded by a write on the same resource);
		\item $XL$ must precede the first write operation for a resource;
	\end{itemize}
	\item a proper unlock must follow an operation:
	\begin{itemize}
		\item $SU$ must follow the last read operation on a resource (skipped if read operation is followed by write on the same resource);
		\item $XU$ must follow the last write operation on a resource;
	\end{itemize}
	\item for every pair of conflicting operations (different transactions, same resource, at least one write), the lock release for the operation occurring earlier must precede the lock acquisition for the operation occurring later;
	\item all locks must precede all unlocks within the same transaction.
\end{itemize}

Each of the previous conditions is expressed as an inequality where each side is either a time point (at which a triple in the schedule occurs) or a lock/unlock request made by some transaction on a resource. We write $SL_i^x$ to indicate a shared lock request by transaction $i$ on resource $x$ and we sometimes write $SL_i^{x[t]}$ to indicate that the lock refers to the operation occurring at time $t$; we use a similar notation for $XL$, $SU$, $XU$.
Additionally, we add inequalities for ordering the time points involved in the schedule ($1<2<\ldots<n$, where $n$ is the number of triples in the schedule).

Let $\I(S)$ be the set of inequalities deriving from the previous conditions on a schedule $S$. Set $\I(S)$ can be exploited to decide 2PL membership: a schedule $S$ is in 2PL if and only if $\I(S)$ is satisfiable.
A possible way to explain non-membership is to derive a minimal subset of $\I(S)$ that is unsatisfiable. However, the cyclicity that ensues may go past the last time point and start over from the beginning, thereby providing a representation defying human intuition.

\begin{example}\label{ex:unsat}
	Consider the following schedule:
 
 $$S_1 =\;r_{1}(y)\;r_{2}(z)\;w_{2}(z)\;r_{1}(x)\;w_{2}(y)\;r_{2}(x)\;w_{2}(x)\;r_{1}(z)$$
The initial system $\I(S_1)$ derived from the requirements of 2PL on $S_1$ consists of 48 inequalities. A minimal subset is obtained by removing inequalities while preserving unsatisfiability of the system. One such minimal set is, e.g., as follows:
$$\left\{\begin{array}{ll}
XU_{2}^{z[3]}<SL_{1}^{z[8]}& \mbox{\footnotesize(lock release must precede lock acquisition for resource $z$ as of $w_{2}^{[3]}(z)$ and $r_{1}^{[8]}(z)$)}\\SL_{1}^{z[8]}<SU_{1}^{x[4]}& \mbox{\footnotesize(lock must precede unlock for the same transaction)}\\SU_{1}^{x[4]}<XL_{2}^{x[7]}& \mbox{\footnotesize(lock release must precede lock acquisition for resource $x$ as of $r_{1}^{[4]}(x)$ and $w_{2}^{[7]}(x)$)}\\XL_{2}^{x[7]}<XU_{2}^{z[3]}& \mbox{\footnotesize(lock must precede unlock for the same transaction)}\\\end{array}\right.$$

The inequalities clearly show a cycle in the $<$ relationship, which entails unsatisfiability of the system, which, in turn, entails that $S_1$ is not a 2PL schedule.
However, the reasons for the violation of the policy are hidden in the inequalities and the actual ``culprits'' remain undetected.
\end{example}

A useful, alternative representation of the set of inequalities $\I(S)$ is that of a graph $\G(S)$, in which the nodes are the left- or right-hand sides of the inequalities and there is an arc $(a,b)$ whenever there is an inequality $a<b$.
At this point, satisfiability of $\I(S)$ is equivalent to acyclicity of $\G(S)$, so the schedule $S$ is in 2PL if and only if $\G(S)$ is acyclic.

To this end, it is convenient to try to derive a topological sort of $\G(S)$ (which exists iff $\G(S)$ is acyclic).
If a topological sort exists, this can be linearized into an ordering of the lock/unlock requests that is compatible with the 2PL policy.

\begin{example}\label{ex:sat}
	Consider the schedule
    $$S_2 =\; r_{4}(x)\;w_{3}(x)\;r_{4}(z)\;w_{4}(y)\;r_{2}(x)\;r_{1}(x)\;w_{2}(z)\;w_{3}(y)\;r_{2}(y)\;w_{1}(x)\;w_{1}(y)$$
	
In this case $\I(S_2)$ consists of 72 inequalities. Once converted to a graph, they admit a topological sort, which induces a sequence of lock/unlock requests and time points that is compatible with the 2PL policy.
In particular, in the following representation, items within the same group enclosed by curly brackets are \emph{unordered} and can be permuted in any possible way, while the arrow ($\rightarrow$) indicates that the requests or time points on the left must precede those on the right.
	$\{ \, SL_{4}^{x[1]} \, \} \rightarrow \{ \, SL_{4}^{z[3]}, \,XL_{4}^{y[4]}, \,1 \, \} \rightarrow \{ \, SU_{4}^{x[1]} \, \} \rightarrow \{ \, XL_{3}^{x[2]} \, \} \rightarrow \{ \, 2 \, \} \rightarrow \{ \, 3 \, \} \rightarrow \{ \, SU_{4}^{z[3]}, \,4 \, \} \rightarrow \{ \, XU_{4}^{y[4]}, \,XL_{2}^{z[7]} \, \} \rightarrow \{ \, XL_{3}^{y[8]} \, \} \rightarrow \{ \, XU_{3}^{x[2]} \, \} \rightarrow \{ \, SL_{2}^{x[5]} \, \} \rightarrow \{ \, 5, \,SL_{1}^{x[6]} \, \} \rightarrow \{ \, 6 \, \} \rightarrow \{ \, 7 \, \} \rightarrow \{ \, 8 \, \} \rightarrow \{ \, XU_{3}^{y[8]} \, \} \rightarrow \{ \, SL_{2}^{y[9]} \, \} \rightarrow \{ \, SU_{2}^{x[5]}, \,9, \,XU_{2}^{z[7]} \, \} \rightarrow \{ \, XL_{1}^{x[10]}, \,SU_{2}^{y[9]} \, \} \rightarrow \{ \, XL_{1}^{y[11]}, \,10 \, \} \rightarrow \{ \, 11, \,XU_{1}^{x[10]} \, \} \rightarrow \{ \, XU_{1}^{y[11]} \, \}.$
\end{example}

We observe that the sequence shown in Example~\ref{ex:sat} is just one of the possible linearizations and that many more pairs/groups are unordered. This will allow a relocation of sequence items that might make it more readable.

In particular, we advocate a representation of the 2PL analysis in tabular form, with a clear indication of when the lock/unlock requests take place and which of them cause violation of the policy. This would provide a more intuitive visualization of both violation and satisfaction of the 2PL policy, as the next examples show.
An additional, useful indication is the \emph{plateau} point of transactions, shown as a vertical, dashed, gray line with the transaction number at the bottom. Such a point corresponds to the moment at which all locks have been acquired and no lock has been released yet.
Finally, we observe that, although the positioning of items within the same unordered group is completely arbitrary, in order to better grasp the sequence of events in a schedule, a useful heuristics is to push lock requests as far as possible towards the right end of the sequence (even past their group) and to separate operations from requests on the same resource.

\begin{example}
	Consider again the schedule $S_2$ from Example~\ref{ex:sat}.
A tabular representation induced by the topological sort is as follows, where colored arrows indicate lock/unlock requests and the first row in the table positions the time points of the various read/write operations in the schedule.

\begin{minipage}[t]{.7\textwidth}
\vspace{0pt}
	\resizebox{\ifdim\textwidth>\width\width\else\textwidth\fi}{!}{\SetTblrInner{colsep=0pt}
	\begin{tblr}{cells ={c},
	column{1}={5mm},rows={6mm},vline{2}={black},
	hlines={black},hline{Z}={0pt},
	vline{11}={gray,dashed},
	vline{11} = {Z}{text=\clap{3}},
	vline{23}={gray,dashed},
	vline{23} = {Z}{text=\clap{1}},
	vline{20}={gray,dashed},
	vline{20} = {Z}{text=\clap{2}},
	vline{4}={gray,dashed},
	vline{4} = {Z}{text=\clap{4}}
	}& &1   & & &2 &3 &4  & & & & &5 & &6  &7 &8 & & &9   &  &10  &11  & &\\
	$x$&\SLarrow{4}&$r_{4}$&\SUarrow{4}&\XLarrow{3}&$w_{3}$&&&&&\XUarrow{3}&\SLarrow{2}&$r_{2}$&\SLarrow{1}&$r_{1}$&&&&&\SUarrow{2}&\gradedXLarrow{1}&$w_{1}$&\XUarrow{1}&&\\
	$y$&&\XLarrow{4}&&&&&$w_{4}$&\XUarrow{4}&\XLarrow{3}&&&&&&&$w_{3}$&\XUarrow{3}&\SLarrow{2}&$r_{2}$&\SUarrow{2}&\XLarrow{1}&$w_{1}$&\XUarrow{1}&\\
	$z$&&\SLarrow{4}&&&&$r_{4}$&\SUarrow{4}&&&&&&&\XLarrow{2}&$w_{2}$&&&&\XUarrow{2}&&&&&\\
	&
	\end{tblr}}
	\\
	\\
\end{minipage}
\begin{minipage}[t]{.3\textwidth}
\vspace{0pt}
	\noindent\textit{Legend}\\
	\SLarrow{}: read lock request\\
	\XLarrow{}: write lock request\\
	\gradedXLarrow{}: lock upgrade\\
	\XUarrow{}: unlock request
\end{minipage}

Examples of the application of the heuristics for positioning/separating requests are:
\begin{itemize}
	\item the lock request $XL_2^{z[7]}$ is pushed as far as possible in the schedule, from its initial position right after time $4$ to right before time $7$, when its corresponding operation takes place;
	\item the group $\{ \, 5, \,SL_{1}^{x[6]} \, \}$ is separated in the resulting table, since the operation at time $5$ is a read on $x$ and $SL_{1}^{x[6]}$ is also on $x$.
\end{itemize}
\end{example}

Yet, the benefits of a tabular representation are even more evident in case of non-2PL schedules, as it the main causes of conflict can be highlighted in a graphical and intuitive way.
However, when the schedule does not comply with 2PL, there is no topological sort that can be exploited for composing the tabular representation.
This requires a repairing action on the graph consisting in the detection of a minimal cycle in the graph, e.g., through breadth-first search (BFS), and in the removal of an item in the cycle so as to try to force satisfiability.
In order to break a cycle, it generally suffices to remove a single inequality. The candidates for removal are preferentially those in which both sides of the inequality refer to the same transaction, the left-hand side is a lock request and the right-hand side an unlock request, with the time of the operation in the lock greater than that of operation in the unlock request. However, there are schedules in which no such inequality can be found and we simply remove one inequality at random in the minimal cycle.

Once an inequality is removed, the resulting graph may have become acyclic. However, there are cases in which the graph remains cyclic after one removal, so further removals are needed. The process goes on with the cyclicity check, the detection of a minimal cycle and the removal of one inequality until the graph becomes acyclic.
At this point, a topological sort of the graph is possible, which, in this case, represents a portion of the schedule that could be executed without violating the 2PL policy. The eliminated inequalities represent those primary causes of violation that we may try to highlight graphically to emphasize the conflicts (to avoid clutter, we choose to only emphasize the first removed inequality). In addition, no transaction with a lock request involved in a discarded inequality will be able to reach the plateau point, so this is removed from the resulting table.

\begin{example}
	Consider again the schedule $S_1$ from Example~\ref{ex:unsat}.
The minimal cycle of inequalities shown there may be broken by removing, e.g., the inequality $SL_1^{z[8]}<SU_1^{x[4]}$, which respects all the desiderata for preferential removal.
However, the resulting system, consisting of one less inequality (47) is still unsatisfiable; a minimal witness of unsatisfiability is the following set:

$$\left\{\begin{array}{ll}
XL_{2}^{y[5]}<XU_{2}^{z[3]}\\
XU_{2}^{z[3]}<SL_{1}^{z[8]}\\
SL_{1}^{z[8]}<SU_{1}^{y[1]}\\
SU_{1}^{y[1]}<XL_{2}^{y[5]}\\\end{array}\right.$$

We can now discard, e.g., $SL_1^{z[8]}<SU_1^{y[1]}$, which is also preferentially removable.
At this point, the remaining set of inequalities is satisfiable and a topological sort can be derived from it, containing all the lock/unlock requests that could be placed compatibly with 2PL.
This topological sort is further enriched with special markers for the two requests corresponding to the first discarded inequality, which can be identified as one of the causes of non-compliance. In the present example, we will mark the conflict between $SL_1^{z[8]}$ and $SU_1^{x[4]}$ with two red circles. In particular, the $SL$ request belongs to transaction $1$, which then cannot reach the plateau point. 

\begin{minipage}[t]{.7\textwidth}
\vspace{0pt}
\resizebox{\ifdim\textwidth>\width\width\else\textwidth\fi}{!}{\SetTblrInner{colsep=0pt}
\begin{tblr}{cells ={c},
column{1}={5mm},rows={6mm},vline{2}={black},
hlines={black},hline{Z}={0pt},
vline{11}={gray,dashed},
vline{11} = {Z}{text=\clap{2}}
}& &1   &2 &  &3 &4  &5   &6 & &7   &  &8 & &\\
$x$&&\SLarrow{1}&&&&$r_{1}$&\SLarrow{2}\redcircled{\SUarrow{1}}&$r_{2}$&\gradedXLarrow{2}&$w_{2}$&\XUarrow{2}&&&\\
$y$&\SLarrow{1}&$r_{1}$&&\SUarrow{1}&&\XLarrow{2}&$w_{2}$&&&\XUarrow{2}&&&&\\
$z$&&\SLarrow{2}&$r_{2}$&\gradedXLarrow{2}&$w_{2}$&&&&&\XUarrow{2}&\redcircled{\SLarrow{1}}&$r_{1}$&\SUarrow{1}&\\
&
\end{tblr}}
\\
\\
\end{minipage}
\begin{minipage}[t]{.3\textwidth}
\vspace{0pt}
\noindent\textit{Legend}\\
\SLarrow{}: read lock request\\
\XLarrow{}: write lock request\\
\gradedXLarrow{}: lock upgrade\\
\XUarrow{}: unlock request
\end{minipage}

\end{example}

Overall, the procedure we follow for deriving the tabular analysis shown in the previous examples is illustrated in Algorithm~\ref{alg:algo}.

\begin{algorithm}[t]
\scalebox{.95}
   {
    \begin{minipage}{1.33\textwidth}
	\begin{enumerate}[itemsep=0pt, parsep=0pt]
			    \item[Input:] \emph{schedule $S$}
			    \item[Output:] \emph{a sequence of (possibly marked) lock/unlock requests and time points}
		\item $\G := \G(S)$
			    \item\label{line:parallel-for} \codewhile\ $\G$ is cyclic
			    \item \quad $M := $ minimal cycle in $\G$ (via BFS)
			    \item \quad $(a,b) := $ arc in $M$ (possibly of the preferred kind)
		\item \quad $\G := \G \setminus \{(a,b)\}$
		\item \quad \codeif\ no marking present \codethen\ mark $a$ and $b$ as ``culprits''
		\item $T := $\texttt{TopSort}$(\G)$
		\item linearize $T$ by
		\item \quad pushing locks in $T$ as much as possible to the right
		\item \quad separating in $T$ operations from requests on the same resource
		\item \codereturn\ $T$
	\end{enumerate}	    
	\end{minipage}
   }
	\caption{Computation of the sequence of 2PL requests in a schedule.}
	\label{alg:algo}
\end{algorithm}

\section{Related Works and Discussion}\label{sec:related}

Database schedules are central to ensuring consistency and isolation in multi-transaction environments. Besides the mentioned 2PL policy, several classes of schedules are widely studied: \textit{view serializable}, \textit{conflict serializable}, and \textit{timestamp-based schedules}.

View Serializability ensures that a schedule produces the same final outcome as a serial schedule by preserving the read and write dependencies. The foundational concepts of view serializability were introduced in~\cite{eswaran1976notions} and further formalized in the theory of database concurrency control~\cite{papadimitriou1979theory}.

A stricter subset of serializability, conflict serializability guarantees that transactions can be reordered without conflicts, such as write-write or read-write conflicts. Gray~\cite{gray1978notes} and Bernstein et al.~\cite{bernstein1987concurrency} provide comprehensive discussions of conflict serializability and its role in ensuring correct schedules.

Using logical timestamps to order transactions, the class of Timestamp-Based Schedules ensures serializability by assigning each transaction a unique timestamp. The concept was influenced by Lamport's seminal work on logical clocks~\cite{lamport1978time} and Reed's development of timestamp-based concurrency control~\cite{reed1978naming}.

These schedule classes remain fundamental to the design of database systems, providing the theoretical foundation for modern concurrency control techniques~\cite{bernstein1987concurrency, silberschatz2020database}.

When transactions are subject to \textit{integrity constraints}~\cite{DBLP:conf/lopstr/ChristiansenM03,DBLP:conf/foiks/ChristiansenM04,DBLP:conf/fqas/Martinenghi04,DBLP:journals/aai/ChristiansenM00,DBLP:conf/dexaw/DeckerM07,DBLP:conf/lpar/DeckerM06,DBLP:conf/dexaw/DeckerM06,DBLP:journals/tplp/CaliM10}—such as foreign key constraints, uniqueness, or domain-specific rules—additional challenges arise. Integrity constraints often introduce dependencies between data items, which complicate the management of locks and increase the potential for deadlocks and performance bottlenecks. For instance, enforcement of foreign key constraints during concurrent updates may require fine-grained locks or careful sequencing of operations~\cite{bernstein1987concurrency}. Similarly, the presence of integrity constraints can amplify the need for predicate locking, as introduced in~\cite{eswaran1976notions}, to ensure that constraints are preserved during concurrent reads and writes. 

Recent works have also explored optimizations to standard 2PL in this context, such as adaptive locking protocols that account for specific constraint patterns, reducing the likelihood of conflicts~\cite{weikum2001transactional}. These studies demonstrate that while traditional locking mechanisms like 2PL are robust, their performance can be significantly impacted by the complexity and frequency of constraint checks. Addressing these challenges remains a critical area of research in modern database systems. A few, dedicated works have studied the interaction between locks and integrity maintenance~\cite{M:PHD2005,DBLP:conf/dexa/MartinenghiC05}, but the issue still needs further investigation.

Integrity constraints are sometimes expressed to indicate intentions or preferences~\cite{Ch03,Kie02,DBLP:conf/er/CiacciaMT19,DBLP:journals/jacm/CiacciaMT20,DBLP:conf/fqas/MartinenghiT09} and, in that respect, they may be violated, thus posing less pressing challenges on the enforcement of concurrency control policies such as 2PL. Violation is also tolerated, although with the tacit aim to repair the data and restore consistency, perhaps after measuring it~\cite{DBLP:journals/jiis/GrantH06,DBLP:conf/er/DeckerM09,DBLP:conf/ijcai/GrantH11,DBLP:conf/ecsqaru/GrantH13,DBLP:journals/ijar/GrantH17,DBLP:journals/ijar/GrantH23}, in paraconsistent or \emph{inconsistency-tolerant} contexts~\cite{DBLP:conf/lpar/DeckerM06,DBLP:conf/dexaw/DeckerM06,DBLP:conf/dexaw/DeckerM07,DBLP:conf/ppdp/DeckerM08}.
In other approaches, such as in the context of ranking, preferences are expressed by means of scoring functions~\cite{DBLP:journals/csur/IlyasBS08,DBLP:journals/pvldb/MartinenghiT10,DBLP:journals/tkde/MartinenghiT12,CM:PACMMOD2024} or by constraints on the weights of scoring functions~\cite{DBLP:journals/tods/CiacciaM20,DBLP:journals/pvldb/CiacciaM17,DBLP:conf/cikm/CiacciaM18,DBLP:conf/sebd/CiacciaM18,DBLP:conf/sisap/BedoCMO19,DBLP:conf/sebd/CiacciaM19,DBLP:conf/sigmod/MouratidisL021} and, therefore, brought to the query level. It should be interesting to assess the interaction of these kinds of queries with the level of concurrency granted by locking policies.

Other kinds of constraints exist, for instance, at the structural level, to govern access to data~\cite{CM:ICDE2008,DBLP:journals/tplp/CaliM10,DBLP:conf/er/CaliM08,CCM:JUCS2009} through, e.g., access patterns. The interaction of these with locking policies is, to the best of our knowledge, still unexplored.

We also note that concurrency control is a ubiquitous concern that involves all kinds of data-intensive applications, such as the typical steps of data pipelines and their core tasks, including Machine Learning algorithms~\cite{DBLP:conf/fqas/Masciari09,DBLP:journals/isci/MasciariMZ14,DBLP:conf/ismis/MasciariMPS20,DBLP:conf/ideas/MasciariMPS20}, and all kinds of data from all sorts of heterogeneous sources, including  RFID~\cite{DBLP:conf/ideas/FazzingaFMF09,DBLP:journals/tods/FazzingaFFM13}, pattern mining~\cite{DBLP:conf/ideas/MasciariGZ13}, crowdsourcing applications~\cite{DBLP:conf/socialcom/GalliFMTN12,DBLP:conf/www/BozzonCCFMT12,DBLP:conf/mmsys/LoniMGGMAMMVL13}, and streaming data~\cite{DBLP:journals/jiis/CostaMM14}.

\section{Conclusions}\label{sec:conclusions}

We have presented a rich set of examples showing an idea for a tool analyzing schedules under the perspective of 2PL membership. While 2PL schedules are fully understood by now, providing support for the automatic analysis of schedules is a less explored topic, which we developed by proposing an algorithm that expresses 2PL constraints as inequalities in graph form, decides membership by trying to detecting cycles in the graph, and breaks such cycles by removing inequalities that might cause cyclicity. When no more cycles exist, the resulting system is used to produce a linearized order of lock and unlock requests that are compatible with the 2PL policy and, if any inequalities were eliminated in the process, these are marked as ``culprits'' for non-membership.
Our examples also show the rendering in tabular form of these linearizations, as produced by an experimental tool we have developed, which generates the results as LaTeX tables.

\bibliographystyle{plain}

\end{document}